# Security Considerations for Remote Electronic Voting over the Internet

Avi Rubin

AT&T Labs – Research
Florham Park, NJ
rubin@research.att.com
http://avirubin.com/

**Abstract**

*This paper discusses the security considerations for remote electronic voting in public elections. In particular, we examine the feasibility of running national federal elections over the Internet. The focus of this paper is on the limitations of the current deployed infrastructure in terms of the security of the hosts and the Internet itself. We conclude that at present, our infrastructure is inadequate for remote Internet voting.*

## 1 Introduction

The right of individuals to vote for our government representatives is at the heart of the democracy that we enjoy. Historically, great effort and care has been taken to ensure that elections are conducted in a fair manner such that the candidate who should win the election based on the vote count actually does. Of equal importance is that public confidence in the election process remain strong. In the past changes to the election process have proceeded deliberately and judiciously, often entailing lengthy debates over even the minutest of details. These changes are approached so sensitively because a discrepancy in the election system threatens the very principles that make our society free, which in turn, affects every aspect of the way we live.

Times are changing. We now live in the Internet era, where decisions cannot be made quickly enough, and there is a perception that anyone who does not jump on the technology bandwagon is going to be left far behind. Businesses are moving online at astonishing speed. The growth of online interaction and presence can be witnessed by the exponential increase in the number of people with home computers and Internet access. There is a prevailing sentiment that any organization that continues in the old ways is obsolete. So, despite the natural inclination to treat our election process as the precious, delicate and fragile process that it is, the question of using the new advances in technology to improve our elections is natural.

The feasibility of *remote electronic voting* in public elections is currently being studied by the National Science Foundation by request of the President of the United States (see http://www.netvoting.org/). Remote electronic voting refers to an election process whereby people can cast their votes over the Internet, most likely through a web browser, from the comfort of their home, or possibly any other location where they can get Internet access. There are many aspects of elections besides security that bring this type of voting into question. The primary ones are

> *coercibility* the danger that outside of a public polling place, a voter could be coerced into voting for a particular candidate.
> *vote selling* the opportunity for voters to sell their vote.
> *vote solicitation* the danger that outside of a public polling place, it is much more difficult to control vote solicitation by political parties at the time of voting.
> *registration* the issue of whether or not to allow online registration, and if so, how to control the level of fraud.

The possibility of widely distributed locations where votes can be cast changes many aspects of our carefully controlled elections as we know them. The relevant issues are of great importance, and could very well influence whether or not such election processes are desirable. However, in this paper, we focus solely on the security considerations as they relate to conducting online public elections. In particular, we look at remote online voting, as opposed to online voter registration, which is a separate, but important and difficult problem. We also focus solely on public elections, as opposed to private elections, where the threats are not as great, and the environment can be more controlled.

The importance of security in elections cannot be overstated. The future of our country, and the free world for that matter, rests on public confidence that the people have the power to elect their own government. Any process that has the potential to threaten the integrity of the system, or even the perceived integrity of the system, should be treated with the utmost caution and suspicion.

## 2   The voting platform

The type of remote electronic voting that we discuss in this paper involves regular Internet users with personal computers and standard operating systems and software. For the sake of the discussion, we focus on Intel machines running Microsoft operating systems with Microsoft or Netscape browsers, and voters participating from home, communicating over a TCP/IP network attached to the Internet. While this is a simplification, it is representative of the vast majority of users under consideration. In this discussion, we refer to the voting platform simply as a *host*.

Threats to hosts can be described as a *malicious payload* and a *delivery mechanism* (A malicious payload is software or configuration information designed to do harm.). Both of these have advanced in sophistication and automation in the past couple of years. The attacks are more sophisticated in the sense that they can do more damage, are more likely to succeed, and disguise themselves better than before. They are more automated in that more and more toolkits have been developed to enable unsophisticated computer users to launch the attacks.

## 2.1 Malicious payload

There are literally hundreds of attack programs that we could discuss in this section. One only need to visit the web site of any number of security software vendors to see the long lists of exploits that affect hosts to various degrees. The fact of the matter is that on the platforms currently in the most widespread use, once a malicious payload reaches a host, there is virtually no limit to the damage it can cause. With today's hardware and software architectures, a malicious payload on a voting client can actually change the voter's vote, without the voter or anyone else noticing, regardless of the kind of encryption or voter authentication in place. This is because the malicious code can do its damage before the encryption and authentication is applied to the data. The malicious module can then erase itself after doing its damage so that there is no evidence to correct, or even detect the fraud. To illustrate, we focus the discussion on two particular malicious payloads that each exemplify the level of vulnerability faced by hosts.

The first program we describe, Backorifice 2000 (BO2K) is packaged and distributed as a legitimate network administration toolkit. In fact, it is very useful as a tool for enhancing security. It is freely available, fully open source, extensible, and stealth (defined below). The package is available at http://www.bo2k.com/. BO2K contains a remote control server that when installed on a machine, enables a remote administrator (or attacker) to view and control every aspect of that machine, as though the person were actually sitting at the console. This is similar in functionality to a commercial product called PCAnywhere. The main differences are that BO2K is available in full source code form and it runs in stealth mode.

The open source nature of BO2K means that an attacker can modify the code and recompile such that the program can evade detection by security defense software (virus and intrusion detection) that look for known *signatures* of programs. A signature is a pattern that identifies a particular known malicious program. The current state of the art in widely deployed systems for detecting malicious code does not go much beyond comparing a program against a list of attack signatures. In fact, most personal computers in peoples' houses have no detection software on them. BO2K is said to run in stealth mode because it was carefully designed to be very difficult to detect. The program does not appear in the Task Menu of running processes, and it was designed so that even an experienced administrator would have a difficult time discovering that it was on a computer. The program is difficult to detect even while it is running.

There can be no expectation that an average Internet user participating in an online election from home could have any hope of detecting the existence of BO2K on his computer. At the same time, this program enables an attacker to watch every aspect of the voting procedure, intercept any action of the user with the potential of modifying it without the user's knowledge, and to further install any other program of the attackers desire, even ones written by the attacker, on the voting user's machine. The package also monitors every keystroke typed on the machine and has an option to remotely lock the keyboard and mouse. It is difficult, and most likely impossible, to conceive of a web application (or any other) that could prevent an attacker who installs BO2K on a user's machine from being able to view and/or change a user's vote.

The second malicious payload that is worth mentioning is the CIH virus, also known as the Chernobyl virus. There are two reasons why we choose this example over the many other possible ones. The first is that the malicious functionality of this virus is triggered to activate on a particular day. April 26, 1999 was a disastrous day in Asia, where the virus had not been that well known, and thousands of computers were affected. This raises concern because election dates are known far in advance. The second reason for choosing this example is that the damage that it caused was so severe, that it often required physically taking the computer to the shop for repair. The code modified the BIOS of the system in such a way that it could not boot. The BIOS is the part of the computer that initializes and manages the relationships and data flow between the system devices, including the hard drive, serial and parallel ports, and the keyboard. A widespread activation of such a virus on the day of an election, or on a day leading up to an election could potentially disenfranchise many voters, as their hosts would not be usable. This threat is increased by the possibility that the spread of the virus could be orchestrated to target a particular demographic group, thus having a direct effect on the election, and bringing the integrity of the entire process into question.

It does not take a very sophisticated malicious payload to disrupt an election. A simple attack illustrates how easy it is to thwart a web application such as voting. Netscape and Internet Explorer, the two most common browsers have an option setting that indicates that all web communication should take place via a *proxy*. A proxy is a program that is interposed between the client and the server. It has the ability to completely control all Internet traffic between the two. Proxies are useful for many Internet applications and for sites that run certain kinds of firewalls. The user sets a proxy by making a change in the preferences menu. The browser then adds a couple of lines to a configuration file. For example, in Netscape, the existence of the following lines in the file

```
c:\program_files\netscape\prefs.js
```

delivers all web content to and from the user's machine to a program listening on port `1799` on the machine `www.badguy.com`.

```
user_pref("network.proxy.http", "www.badguy.com");
user_pref("network.proxy.http_port", 1799);
```

If an attacker can add these two lines (substituting his hostname for `www.badguy.com`) to the preferences file on somebody's machine, he can control every aspect of the web experience of that user. There also ways of doing this without leaving a trail that leads directly to the attacker. While proxies cannot be used to read information in a secure connection, they can be used to spoof a user into a secure connection with the attacker, instead of the actual voting server, without the user realizing it. The next section explains various ways that an attacker could effect changes on a voter's computer.

## 2.2 Delivery mechanism

The previous section gave three examples of what an attacker could do to disrupt an election if the attacker could install code of his choosing on peoples' computers. This section deals with how this installation could happen.

The first, and most obvious mechanism is physical installation. Most people do not keep their computers in a carefully controlled, locked environment. Imagine someone who develops an application to attack the voting system, such as the two described above, prepares a floppy disk with the code on it, and then installs it on as many machines as possible. This could be accomplished by breaking into houses, by accessing machines in someone's house when visiting, by installing the program on public machines in the library, etc. The bottom line is that many people can obtain physical access to many other peoples' computers at some point leading up to an election. Then, malicious code can be delivered that can trigger any action at a later date, enable future access (as in the case of BO2K), or disrupt normal operation at any time. Considering that many of the attack programs that we are seeing these days run in stealth mode, malicious code could be installed such that average computer users cannot detect its presence.

While the physical delivery of malicious code is a serious problem, it is nowhere near as effective as remote automated delivery. By now, most people have heard of the Melissa virus and the I Love You bug. These are the better-known ones, but many such attacks happen all the time. In fact, the most widespread of the e-mail viruses, Happy99, has received very little media attention. Typically, these attacks cause temporary disruption in service, and perform some annoying action. In most of the cases, the attacks spread wider and faster than their creators ever imagined. One thing that all of these attacks have in common is that they install some code on the PCs that are infected. There is a misconception by many people that users must open an attachment in order to activate them. In fact, one virus called Bubbleboy was triggered as soon as a message was previewed in the Outlook mailer, requiring no action on the part of the user. Any one of these e-mail viruses could deliver the attack code described in the previous section.

It is naïve to think that we have seen the worst of the Internet viruses, worms, and bugs. In the last several months, the incidents of new attacks have grown much faster than our ability to cope with them. This is a trend that is likely to continue.

E-mail viruses are not the only way that malicious code can be delivered to hosts. The computers in most peoples' houses are running operating systems with tens of thousands of lines of code. These systems are known to be full of operational bugs as well as security flaws. On top of these platforms, users are typically running many applications with security problems. These security flaws can be exploited remotely to install malicious code on them. The most common example of such a flaw is a buffer overflow. A buffer overflow occurs when a process assigns more data to a memory location than was expected by the programmer. The consequence is that that attacker can manipulate the computer's memory to cause arbitrary malicious code to run. There are ways to check for and prevent this in a program, and yet buffer overflows are the most common form of security flaw in deployed systems today.

Perhaps the most likely candidate for delivering a widespread attack against an election is an ActiveX control, downloaded automatically and unknowingly from a Web server, which installs a Trojan horse (hidden program) that later interferes with voting. Several documented attacks against Windows systems operated exactly this way. In fact, any application that users are lured into downloading can do the same. This includes browser plug-ins, screen savers, calendars, and any other program that is obtained over the Internet. Another danger is that the application itself may be clean, but the installer might install a dynamically linked library (DLL) or other malicious module, or overwrite operating system modules. The number of ways is legion, and most users are not aware of the dangers when they add software to their computers. As long as there are people out there who download and install software over the Internet onto today's personal computers running today's operating systems, it will be easy for attackers to deliver code that changes their votes, to peoples' computers.

User's who open attachments and download software from the network are not the only ones putting their votes at risk. AOL, for instance, is in a position to control a large fraction of the total votes, because all of their users run AOL's proprietary software. There are dozens of software vendors whose products run on many peoples' home machines. For example, there are millions of personal computers running Microsoft office, Adobe Acrobat, RealPlayer, WinZip, Solitaire, and the list goes on. These vendors are in a position to modify any configuration file and install any malicious code on their customers' machines, as are the computer manufacturers and the computer vendors. Even if the company is not interested in subverting an election, all it takes is one rogue programmer who works for any of these companies. Most of the software packages require an installation procedure where the system registry is modified, libraries are installed, and the computer must reboot. During any stage of that process, the installation program has complete control of all of the software on that machine. In current public elections, the polling site undergoes careful scrutiny. Any change to the process is audited carefully, and on election day, representatives from all of the major parties are present to make sure that the integrity of the process is maintained. This is in sharp contrast to holding an election that allows people to cast their votes from a computer full of insecure software that is under the direct control of several dozen software and hardware vendors and run by users who download programs from the Internet, over a network that is known to be vulnerable to total shutdown at any moment.

## 3  The communications infrastructure

A network connection consists of two endpoints and the communication between them. The previous section dealt with one of the endpoints, the user's host. The other endpoint is the elections server. While it is in no way trivial, the technology exists to provide reasonable protection on the servers. This section deals with the communication between the two endpoints.

Cryptography can be used to protect the communication between the user's browser and the elections server. This technology is mature and can be relied upon to ensure the integrity and confidentiality of the network traffic. This section does not deal with the classic security properties of the communications infrastructure; rather, we look at the *availability* of the Internet service, as required by remote electronic voting over the Internet.

Most people are aware of the massive distributed denial of service (DDOS) attack that brought down many of the main portals on the Internet in February, 2000. While these attacks brought the vulnerability of the Internet to denial of service attacks to the mainstream public consciousness, the security community has long been aware of this, and in fact, this attack was nothing compared to what a dedicated and determined adversary could do. The February attack consisted of the installation and execution of publicly available attack scripts. Very little skill was required to launch the attack, and minimal skill was required to install the attack.

The way DDOS works is that a program called a *daemon* is installed on many machines. Any of the delivery mechanisms described above can be used. One other program is installed somewhere called the *master*. These programs are placed anywhere on the Internet, so that there are many, unwitting accomplices to the attack, and the real attacker cannot be traced. The system lies dormant until the attacker decides that it is time to strike. At that point, the attacker sends a signal to the master, using a publicly available tool, indicating a target to attack. The master conveys this information to all of the daemons, who simultaneously flood the target with more Internet traffic than it can handle. The effect is that the target machine is completely disabled.

We experimented in the lab with one of the well known DDOS programs called Tribe Flood Network (TFN), and discovered that the attack is so potent, that even one daemon attacking a Unix workstation disabled it to the point where it had to be rebooted. The target computer was so overwhelmed that we could not even move the cursor with the mouse.

There are tools that can be easily found by anyone with access to the web that automate the process of installing daemons, masters, and the attack signal. People who attack systems with such tools are known as *script kiddies*, and represent a growing number of people. In an election, the adversary is more likely to be someone at least as knowledgeable as the writers of the script kiddy tools, and possibly with the resources of a foreign government.

There are many other ways to target a machine and make it unusable, and it is not too difficult to target a particular set of users, given domain name information that can easily be obtained from the online registries such as Register.com and Network Solutions, or directly from the WHOIS database. The list of examples of attacks goes on and on. A simple one is the *ping of death*, in which a packet can be constructed and split into two fragments. When the target computer assembles the fragments, the result is a message that is too big for the operating system to handle, and the machine crashes. This has been demonstrated in the lab and in the wild, and script kiddy tools exist to launch it.

The danger to Internet voting is that it is possible that during an election, communication on the Internet will stop because attackers cause routers to crash, election servers to get flooded by DDOS, or a large set of hosts, possibly targeted demographicly, to cease to function. In some close campaigns, even an untargeted attack that changes the vote by one percentage point could sway the election.

## 4   Social engineering

*Social Engineering* is the term used to describe attacks that involve fooling people into compromising their security. Talking with election officials, one discovers that one of the issues that they grapple with is the inability of many people to follow simple directions. It is surprising to learn that, for example, when instructed to circle a candidate's name, people will often underline it. While computers would seem to offer the opportunity to provide an interface that is tightly controlled and thus less subject to error, this is counter to the typical experience most users have with computers. For non-Computer Scientists, computers are often intimidating and unfamiliar. User interfaces are often poor and create confusion, rather than simplifying processes.

A remote voting scheme will have some interface. The actual design of that interface is not the subject of this paper, but it is clear that there will be some interface. For the system to be secure, there must be some way for voters to know that they are communicating with the election server. The infrastructure does exist right now for computer security specialists, who are suspicious that they could be communicating with an imposter, to verify that their browser is communicating with a valid election server. The SSL protocol and server side certificates can be used for this. While this process has its own risks and pitfalls, even if we assume that it is flawless, it is unreasonable to assume that average Internet users who want to vote on their computers can be expected to understand the concept of a server certificate, to verify the authenticity of the certificate, and to check the active ciphersuites to ensure that strong encryption is used. In fact, most users would probably not distinguish between a page from an SSL connection to the legitimate server and a non-SSL page from a malicious server that had the exact same look as the real page.

There are several ways that an attacker could spoof the legitimate voting site. One way would be to send an e-mail message to a user telling that user to click on a link, which would then bring up the fake voting site. The adversary could then collect the user's credentials and in a sense, steal the vote. An attacker could also set up a connection to the legitimate server and feed the user a fake web page, and act as a man in the middle, transferring information between the user and the web server, with all of the traffic under the attacker's control. This is probably enough to change a user's vote, regardless of how the application is implemented.

A more serious attack is possible by targeting the Internet's Domain Name Service (DNS). The DNS is used to maintain a mapping from IP addresses, which computers use to reference each other (e.g. 135.207.18.199) to domain names, which people use to reference computers (e.g. [www.research.att.com](www.research.att.com)). The DNS is known to be vulnerable to attacks, such as cache poisoning, which change the information available to hosts about the IP addresses of computers. The reason that this is serious is that a DNS cache poisoning attack, along with many other known attacks against DNS, could be used to direct a user to the wrong web server when the user types in the name of the election server in the browser. Thus, a user could follow the instructions for voting, and yet receive a page that looked exactly like what it is supposed to look like, but actually is entirely controlled by the adversary. Detailed instructions about checking certificate validity are not likely to be understood nor followed by a substantial number of users.

Another problem along these lines is that any computer under the control of an adversary can be made to simulate a valid connection to an election server, without actually connecting to anything. So, for example, a malicious librarian or cyber café operator could set up public computers that appear to accept votes, but actually do nothing with the votes. This could even work if the computers were not connected to the Internet, since no messages need to be sent or received to fool a user into believing that their vote was cast. Setting up such machines in districts known to vote a certain way could influence the outcome of an election.

## 5 Specialized devices

One potential enabler at our disposal is the existence of tamper-resistant devices, such as smart cards. Cryptographic keys can be generated and stored on these devices, and they can perform computations, such that proper credentials can be exchanged between a client and a voting server. However, there are some limitations to the utility of such devices. The first is that there is not a deployed base of smart card readers on peoples' personal computers. Any system that involves financial investment on the part of individuals in order to vote is unacceptable. Some people are more limited in their ability to spend, and it is unfair to decrease the likelihood that such people vote. It would, in effect, be a poll tax. This issue is often referred to as the *digital divide*.

Even if everybody did have smart card readers on their computers, there are security concerns. The smart card does not interact directly with the election server. The communication goes through the computer. Malicious code installed on the computer could misuse the smart card. At the very least, the code could prevent the vote from actually being cast, while fooling the user into believing that it was. At worst, it could change the vote.

Other specialized devices, such as a cell phone with no general-purpose processor, equipped with a smart card, offer more promise of solving the technical security problems. However, they introduce even greater digital divide issues. In addition, the user interface issues, which are fundamental to a fair election, are much more difficult. This is due to the more limited displays and input devices. Finally, while computers offer some hope of improving the accessibility of voting for the disabled, specialized devices are even more limiting in that respect.

# 6 Is there hope?

Given the current state of insecurity of hosts and the vulnerability of the Internet to manipulation and denial of service attacks, there is no way that a public election of any significance involving remote electronic voting could be carried out securely. So, is there any hope that this will change?

For this to happen, the next generation of personal computers that are widely adopted must have hardware support to enable a *trusted path* between the user and the election server. There must be no way for malicious code to be able to interfere with the normal operation of applications. Efforts such as the Trusted Computing Platform Alliance (TCPA) (see http://www.trustedpc.org/home/home.htm) must be endorsed. The challenge is great because to enable secure remote electronic voting, the vast majority of computer systems need to have the kind of high assurance aspired to by the TCPA. It is not clear whether or not the majority of PC manufacturers will buy into the concept. The market will decide. While it is unlikely that remote electronic voting will be the driving force for the design of future personal computers, the potential for eliminating the hazards of online electronic commerce could potentially fill that role.

One reason that remote electronic voting presents such a security challenge is that any successful attack would be very high profile, a factor that motivates much of the hacking activity to date. Even scarier is that the most serious attacks would come from someone motivated by the ability to change the outcome without anyone noticing. The adversaries to an election system are not teenagers in garages but foreign governments and powerful interests at home and abroad. Never before have the stakes been so high.

# 7 Conclusions

A certain amount of fraud exists in the current offline election system. It is tolerated because there is no alternative. The system is localized so that it is very unlikely that a successful fraud could propagate beyond a particular district. Public perception is that the system works, although there may be a few kinks in it here and there. There is no doubt that the introduction of something like remote electronic voting will, and should, come under careful scrutiny, and in fact, the system may be held up to a higher standard. Given the current state of widely deployed computers in peoples' homes, the vulnerability of the Internet to denial of service attacks, and the unreliability of the Domain Name Service, we believe that the technology does not yet exist to enable remote electronic voting in public elections.

# Acknowledgements

We thank all of the participants of the Internet Policy Institute e-voting workshop for a wonderful exchange of ideas. Special thanks go to Lorrie Cranor, Andrew Hume, and David Jefferson for valuable input.